\documentclass[
twocolumn, 
nofootinbib,
secnumarabic,amssymb, preprintnumbers, superscriptaddress, aps, prd]{revtex4-1}
\usepackage{amsmath,times,amssymb,latexsym,graphics}
\usepackage{ascmac,fancybox}
\usepackage{bbm}
\usepackage{braket}
\usepackage{dcolumn}
\usepackage{simplewick}

\newcommand{\be}{\begin{equation}}
\newcommand{\ee}{\end{equation}}
\newcommand{\nn}{\nonumber}

\newcommand{\e}{\epsilon}

\newcommand{\G}{\Gamma}

\newcommand{\p}{\partial}

\newcommand{\bpm}{\begin{pmatrix}}
\newcommand{\epm}{\end{pmatrix}}
\newcommand{\bbm}{\begin{bmatrix}}
\newcommand{\ebm}{\end{bmatrix}}

\begin{document}
\preprint{OU-HET 979}
\title{Entanglement Wedge Cross Section from the Dual Density Matrix}
\date{\today}
\author{Kotaro Tamaoka}\email[]{kotaro.tamaoka@yukawa.kyoto-u.ac.jp}
\affiliation{\it Center for Gravitational Physics,Yukawa Institute for Theoretical Physics, Kyoto University, Kyoto 606-8502, JAPAN and}
\affiliation{\it Department of Physics, Osaka University, Toyonaka, Osaka 560-0043, JAPAN}

\begin{abstract}
We define a new information theoretic quantity called odd entanglement entropy (OEE) which enables us to compute the entanglement wedge cross section in holographic CFTs. The entanglement wedge cross section has been introduced as a minimal cross section of the entanglement wedge, a natural generalization of the Ryu-Takayanagi surface. By using the replica trick, we explicitly compute the OEE for two-dimensional holographic CFT (AdS${}_3$ and planar BTZ blackhole) and see agreement with the entanglement wedge cross section. We conjecture this relation will hold in general dimensions. 
\end{abstract}
\maketitle
\noindent
\section{Introduction and Summary}
The entanglement entropy (EE) quantifies the quantum entanglement between two subsystems for a given pure state. It is defined by the von Neumann entropy of a reduced density matrix $\rho_A$ on a subsystem $A$, $S(\rho_A)=-\textrm{Tr}_{\mathcal{H}_{A}}\rho_{A}\log\rho_{A}$. 
If one considers states in the conformal field theories (CFTs) with the gravity dual\cite{Maldacena:1997re}, the EE tells us the area of the minimal surface anchored on the boundary of asymptotically AdS\cite{Ryu:2006bv, Hubeny:2007xt}. This strongly suggests that the bulk gravity is encoded into the structure of quantum entanglement in the boundary. Since the EE cannot tell us all the structure of the entanglement, the minimal surface also cannot tell us the whole structure of geometry in the same manner. Therefore, finding the generalization on both sides is quite important in order to decode the profound connection between the entanglement and the geometry\cite{Maldacena:2001kr,Swingle:2009bg,VanRaamsdonk:2010pw}.   

Recently, the entanglement wedge cross section $E_W$, a generalization of the minimal surface, has been introduced\cite{Takayanagi:2017knl, Nguyen:2017yqw}. The definition of the minimal surface and the $E_W$ will be reviewed in section \ref{sec:review}. The cross section has been conjectured to be dual to the entanglement of purification (EoP) \cite{Terhal:2002}, which is a correlation measure for mixed states (for recent progress, refer to \cite{Bhattacharyya:2018sbw,Bao:2017nhh,Hirai:2018jwy,Umemoto:2018jpc,Espindola:2018ozt,Umemoto:2018jpc,Bao:2018gck}). The EoP is also a generalization of the EE and has many nice properties consistent with the entanglement wedge cross section. However, computing the EoP is a really hard task because we need to find the minimized value from all possible purifications.

Can we then extract the entanglement wedge cross section {\it directly} from a given mixed state in QFT?
In this letter, we answer yes to this question and demonstrate it explicitly; however, from another (rather ``odd'') generalization of the EE. 

We first summarize the main result of the present letter. Let $\rho_{A_1A_2}$ be a mixed state acting on bipartite Hilbert space $\mathcal{H}=\mathcal{H}_{A_1}\otimes\mathcal{H}_{A_2}$. Then we define a quantity
\be
S^{(n_o)}_{o}(\rho_{A_1A_2})\equiv\dfrac{1}{1- n_o}\left[\textrm{Tr}_{\mathcal{H}}(\rho^{T_{A_2}}_{A_1A_2})^{n_o}-1\right], \label{eq:oeen}
\ee
where $T_{A_2}$ is the partial transposition\cite{Peres:1996dw} with respect to the subsystem $A_2$. Namely, we will consider the Tsallis entropy\cite{Tsallis:1987eu} for the partially transposed $\rho_{A_1A_2}$. We are especially interested in the limit $n_o\rightarrow1$, 
\be
S_{o}(\rho_{A_1A_2})\equiv\lim_{n_o\rightarrow1}S^{(n_o)}_{o}(\rho_{A_1A_2}), \label{eq:OEE}
\ee
where $n_o$ is analytic continuation of an {\it odd} integer\footnote{We should keep in mind that it is not enough to fix a unique analytic continuation as like other quantities computed from the replica trick.}. Since the odd integer analytic continuation is crucial in the later discussion, we will call $S_{o}$ as ``odd entanglement entropy'' or OEE in short. Loosely speaking, the OEE is the von Neumann entropy with respect to $\rho^{T_{A_2}}_{A_1A_2}$; however, $\rho^{T_{A_2}}_{A_1A_2}$ potentially contains negative eigenvalues. 
In section \ref{sec:2}, we will be more precise on that point. In particular, we will demonstrate the following three facts: First, $S_{o}(\rho_{A_1A_2})$ reduces to the EE $S(\rho_{A_1})$ if $\rho_{A_1A_2}$ is a pure state. Second, $S_{o}(\rho_{A_1A_2})$ reduces to the von Neumann entropy $S(\rho_{A_1A_2})$ if $\rho_{A_1A_2}$ is a product state. Third, if one considers two-dimensional holographic CFT, direct calculation indeed agrees with
\be
\mathcal{E}_W(\rho_{A_1A_2})\equiv S_{o}(\rho_{A_1A_2})-S(\rho_{A_1A_2})=E_W(\rho_{A_1A_2}). 
\ee
In particular, we consider the subregion of the vacuum state (section \ref{sec:3}) and the thermal state (section \ref{sec:4}). 
We conjecture this relation will hold even for the higher dimensional cases.  From our viewpoint, the $\mathcal{E}_W$ is similar to the coherent information\cite{Schumacher:1996dy,Horodecki:2005ehk}, which can take even a negative value. We conclude with a discussion on this point in section \ref{sec:4}. 

\section{Entanglement wedge cross section}\label{sec:review}
In this section, we briefly review the holographic prescription of the EE and definition of the entanglement wedge cross section. For the rigorous definition, refer to \cite{Takayanagi:2017knl}. Throughout this letter, we assume a static geometry in the bulk and take a conventional time slice $M$. 

\begin{figure}[htbp]
\begin{center}
\resizebox{85mm}{!}{\includegraphics{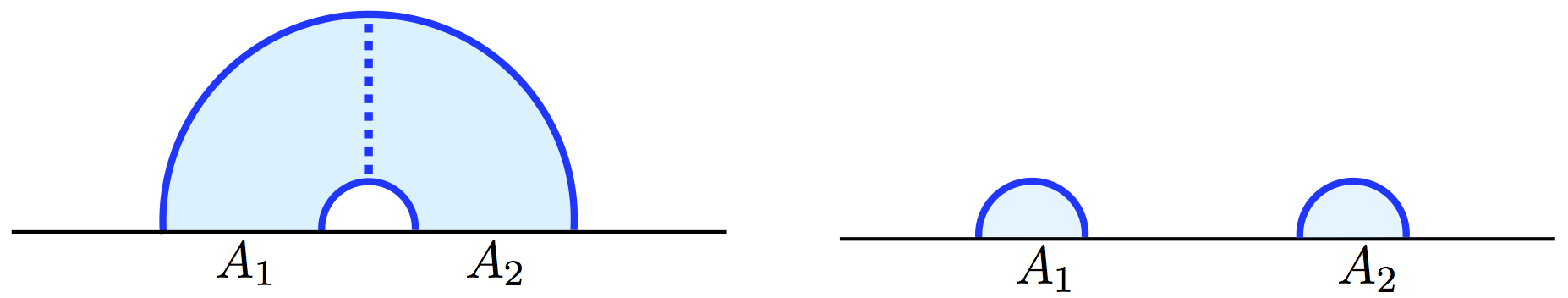}}
\caption{Left: An example of the entanglement wedge cross section $\Sigma^{\textrm{min.}}_{A_1A_2}$ (a blue dotted line). The vertical direction corresponds to the radial one of Poincare AdS${}_3$. The horizontal line coincides with a time slice of CFT${}_2$. Blue curved lines show the minimal surfaces $\G^{\textrm{min.}}_A\, (A=A_1A_2)$ for $S(\rho_{A_1A_2})$ where the $\rho_{A_1A_2}$ is a state acting on bipartite Hilbert space $\mathcal{H}_{A_1}\otimes\mathcal{H}_{A_2}$ (associated with geometrical subregions $A_1$ and $A_2$) and is supposed to be dual to the entanglement wedge. A blue shaded region represents a time slice of the entanglement wedge. Right: If $A_1$ and $A_2$ are sufficiently distant, we have no connected entanglement wedge and  $E_W(\rho_{A_1A_2})=0$.}\label{fig:ewcs_single} 
\end{center}
\end{figure}

To this end, we first recall the holographic EE for the static geometries\cite{Ryu:2006bv}. Let us consider the subregion $A$ in $\p M$. We can imagine a series of co-dimension 2 surfaces $\G_A$ which satisfy $\p\G_A=\p A$ and are homologous to $A$. Then, a minimal area one $\G^{\textrm{min.}}_A$ is called the minimal surface. The holographic EE is given by 
\be
S(\rho_A)=\dfrac{(\textrm{Area of $\G^{\textrm{min.}}_A$})}{4G_N},
\ee
where $G_N$ is the Newtonian constant. Here $\rho_A$ is a reduced density matrix on the subregion $A$. The $\rho_A$ is supposed to be dual to a bulk subregion called the entanglement wedge\cite{Czech:2012bh,Wall:2012uf,Headrick:2014cta}. We define the (time slice of) entanglement wedge as a bulk region surrounded by $\G^{\textrm{min.}}_A$ and $A$\footnote{On the definition of the entanglement wedge, we should include the domain of dependence of the region surrounded by $\G^{\textrm{min.}}_A$ and $A$. Since we will focus on static geometries, it is enough to consider its time slice.}.
Note that we can start even from disconnected subregions in the boundary. For later use, we further divide $A=A_1\cup A_2\equiv A_1 A_2$. See FIG. \ref{fig:ewcs_single}, for a nontrivial example.

Next, we define the entanglement wedge cross section. Let us regard the boundary of entanglement wedge $A\cup \G^{\textrm{min.}}_A$ as a new boundary of the bulk geometry. Then, we can find a new minimal area surface $\Sigma^{\textrm{min.}}_{A_1A_2}$ which separates $A_1$ and $A_2$. An important point is that we let the $\Sigma^{\textrm{min.}}_{A_1A_2}$ end not only on $A$ but also on the $\G^{\textrm{min.}}_A$. 
Since the $\Sigma^{\textrm{min.}}_{A_1A_2}$ can be regarded as a minimal cross section of the entanglement wedge, we define the entanglement wedge cross section as 
\be
E_W(\rho_{A_1A_2})=\dfrac{(\textrm{Area of $\Sigma^{\textrm{min.}}_{A_1A_2}$})}{4G_N}.
\ee
As an example, see again FIG. \ref{fig:ewcs_single}.

\section{Definition and Properties of $S_{o}$ \& $\mathcal{E}_W$}\label{sec:2}
\subsection{Partial transposition and a generalized EE}
We first introduce the partial transposition that is relevant to the definition of \eqref{eq:oeen}. Let $\rho_{AB}$ be a state acting on Hilbert space $\mathcal{H}=\mathcal{H}_A\otimes\mathcal{H}_B$ and let $\ket{e^{(A,B)}_i}$s $(i=1,2,\cdots,\textrm{dim}\mathcal{H}_{A,B})$ be a complete set thereof. Using this basis, we can expand a given density matrix,
\begin{align}
\rho_{AB}&=\sum_{ik}\sum_{j\ell}\braket{e^{(A)}_ie^{(B)}_j|\rho_{AB}|e^{(A)}_ke^{(B)}_\ell}\ket{e^{(A)}_ie^{(B)}_j}\bra{e^{(A)}_ke^{(B)}_\ell}.
\end{align}
We define the partial transposition of the $\rho_{AB}$ with respect to $\mathcal{H}_{A,B}$ as 
\begin{align}
\braket{e^{(A)}_ie^{(B)}_j|\rho^{T_A}_{AB}|e^{(A)}_ke^{(B)}_\ell}&=\braket{e^{(A)}_ke^{(B)}_j|\rho_{AB}|e^{(A)}_ie^{(B)}_\ell},\\
\braket{e^{(A)}_ie^{(B)}_j|\rho^{T_B}_{AB}|e^{(A)}_ke^{(B)}_\ell}&=\braket{e^{(A)}_ie^{(B)}_\ell|\rho_{AB}|e^{(A)}_ke^{(B)}_j}.
\end{align}
Note that the partial transposition does not change its normalization $\textrm{Tr}_{\mathcal{H}}\rho^{T_A}_{AB}=\textrm{Tr}_{\mathcal{H}}\rho^{T_B}_{AB}=\textrm{Tr}_{\mathcal{H}}\rho_{AB}=1$, whereas it changes the eigenvalues. Since the partial transposition is not a completely positive map, the $\rho^{T_B}_{AB}$ can include negative eigenvalues. This negative property is actually a sign of the quantum entanglement\cite{Peres:1996dw} and utilized to, for example, the negativity\cite{Vidal:2002zz}. See also a recent argument on the entanglement wedge cross section and the negativity\cite{Kudler-Flam:2018qjo}.

The $n$-th power of the $\rho^{T_B}_{AB}$ depends on the parity of $n$:
\be
\textrm{Tr}_{\mathcal{H}}(\rho^{T_B}_{AB})^{n}=\left\{
\begin{array}{ll}
\sum_{\lambda_i>0}|\lambda_i|^{n}-\sum_{\lambda_j<0}|\lambda_j|^{n} & (n: \textrm{odd}),\\
\sum_{\lambda_i>0}|\lambda_i|^{n}+\sum_{\lambda_j<0}|\lambda_j|^{n} & (n: \textrm{even}),
\end{array}\right.
\ee
where $\lambda_i$s are the eigenvalues of the $\rho^{T_B}_{AB}$. This argument is completely the same as the negativity using the replica trick\cite{Calabrese:2012ew,Calabrese:2012nk}. The main difference in the present letter is that we are just choosing the odd integer. Therefore, OEE can be formally written as
\be
S_{o}(\rho_{AB})=-\sum_{\lambda_i>0}|\lambda_i|\log|\lambda_i|+\sum_{\lambda_j<0}|\lambda_j|\log|\lambda_j|. \label{eq:oee}
\ee
\subsection{Pure states}
Let $\ket{\Psi_{AB}}$ be a pure state in bipartite Hilbert space $\mathcal{H}=\mathcal{H}_A\otimes\mathcal{H}_B$. Using the Schmidt decomposition, we can write the $\ket{\Psi_{AB}}$ as a simple form,
\be
\ket{\Psi_{AB}}=\sum^N_{n=1}\sqrt{p_n}\ket{n_A}\ket{n_B},
\ee
where $0\leq p_n\leq 1$, $\sum_np_n=1$. The $N$ can be taken as $\min(\dim\mathcal{H}_A,\dim\mathcal{H}_B)$. One can show that the corresponding density matrix $\rho_{AB}=\ket{\Psi_{AB}}\hspace{-1mm}\bra{\Psi_{AB}}$ and its partial transposition $\rho^{T_B}_{AB}$ have the eigenvalues,
\begin{align}
\textrm{Spec}(\rho_{AB})&=\{1,0,\cdots,0\}, \\
\textrm{Spec}(\rho^{T_B}_{AB})&=\{p_1,\cdots,p_N, +\sqrt{p_1p_2},-\sqrt{p_1p_2},\nn\\
&\cdots,+\sqrt{p_{N-1}p_{N}},-\sqrt{p_{N-1}p_{N}}\}. \label{eq:spec_rhot}
\end{align}
Here each $\pm\sqrt{p_ip_j} \,(i\neq j)$ in \eqref{eq:spec_rhot} appears just once respectively. 
In particular, from the definition of the \eqref{eq:oee}, these contributions completely cancel out. Thus, one can conclude that
\be
\mathcal{E}_W(\rho_{AB})=S_{o}(\rho_{AB})=S(\rho_A)\;\; (\textrm{for pure states}).
\ee
\subsection{Product states}
Let $\rho_{A_1B_1}\otimes\sigma_{A_2B_2}$ be a product state with respect to the bipartition $\mathcal{H}_{A_1B_1}\otimes\mathcal{H}_{B_2A_2}$. Then the $S_o$ is {\it additive},
\begin{align}
S_o(\rho_{A_1B_1}\otimes\sigma_{A_2B_2})&=S_o(\rho_{A_1B_1})+S_o(\sigma_{A_2B_2}),
\end{align}
so is $\mathcal{E}_W$. Here we took the partial transposition with respect to $B_1$ and $B_2$.
In particular, if
\be
\tau_{AB}=\tau^\prime_{A}\otimes\tau^{\prime\prime}_{B}, 
\ee
we have $\textrm{Tr}_{\mathcal{H}}\tau_{AB}^n=\textrm{Tr}_{\mathcal{H}}(\tau_{AB}^{T_B})^n$. This fact immediately leads $S_{o}(\tau_{AB})=S(\tau_{AB})$. Thus, we also obtain $\mathcal{E}_W(\tau_{AB})=0$. Note that all properties discussed the above are consistent with the entanglement wedge cross section.
\section{Vacuum state in holographic CFT${}_2$}\label{sec:3}
In this section, we compute the $S_{o}$ and the $\mathcal{E}_W$ for mixed states in CFT on $\mathbb{R}^{2}$. We divide the total Hilbert space of CFT into $\mathcal{H}_A\otimes\mathcal{H}_{A^c}$, where the corresponding subregion $A$ and its complement $A^c$ are not necessarily to be connected. Then  we can prepare a mixed state $\rho_{A_1A_2}\equiv\textrm{Tr}_{\mathcal{H}_{A^c}}\ket{0}\hspace{-1mm}\bra{0}$, where $\ket{0}$ is the vacuum state in CFT. Here we further divided the remaining subspace $\mathcal{H}_{A}$ into two pieces, $\mathcal{H}_{A_1}$ and $\mathcal{H}_{A_2}$. We will focus on the holographic CFT${}_2$. 

\subsection{Two disjoint intervals}
First, we consider disjoint interval $A_1=[u_1,v_1],A_2=[u_2,v_2]$ on a time slice $\tau=0$. In order to compute the $S_{o}$ and the $\mathcal{E}_W$, we can apply the replica trick as usual\cite{Calabrese:2012ew,Calabrese:2012nk}. In particular, one can write the $n$-th power of the density matrix and its partial transposition in terms of the correlation functions for a cyclic orbifold theory $\textrm{CFT}^n/\mathbb{Z}_n$,
\begin{align}
\textrm{Tr}_{\mathcal{H}_{A}}(\rho_{A_1A_2})^{n}&=\braket{\sigma_n(u_1)\bar{\sigma}_n(v_1)\sigma_n(u_2)\bar{\sigma}_n(v_2)}_{\textrm{CFT}^n/\mathbb{Z}_n},\label{eq:rhon}\\
\textrm{Tr}_{\mathcal{H}_{A}}(\rho^{T_{A_2}}_{A_1A_2})^{n}&=\braket{\sigma_n(u_1)\bar{\sigma}_n(v_1)\bar{\sigma}_n(u_2)\sigma_n(v_2)}_{\textrm{CFT}^n/\mathbb{Z}_n},\label{eq:rhotn}
\end{align}
where $\sigma_n (\bar{\sigma}_n)$ is the (anti-)twist operator with scaling dimension $h_{\sigma_n}=\bar{h}_{\sigma_n}=\frac{c}{24}(n-\frac{1}{n})$. In terms of the original $n$-fold geometry, these operators map the $n$-th replica sheet to $n\pm1$-th ones. For later use, we also introduce $\sigma^2_n$ and $\bar{\sigma}^2_n$ which map the $n$-th replica sheet to $n\pm2$-th ones. The scaling dimension of the $\sigma^2_n$ depends on the parity of $n$\cite{Calabrese:2012ew,Calabrese:2012nk},
\be
h_{\sigma^2_n}=\bar{h}_{\sigma^2_n}=\left\{
\begin{array}{ll}
\dfrac{c}{24}\left(n-\dfrac{1}{n}\right) & (n: \textrm{odd}),\\
\dfrac{c}{12}\left(\dfrac{n}{2}-\dfrac{2}{n}\right) & (n: \textrm{even}).
\end{array}\right.
\ee
Since we are interested in the odd integer case, this coincides with $h_{\sigma_n}$.  
Hereafter, we will omit the suffix of the correlation function, $\textrm{CFT}^n/\mathbb{Z}_n$, for brevity. Since \eqref{eq:rhon} is studied in \cite{Hartman:2013mia}, we focus on the latter one. 

Let us expand \eqref{eq:rhotn} into the conformal blocks in t-channel, 
\begin{align}
&\bcontraction{\langle\sigma_n(u_1)}{\bar{\sigma}_n}{(v_1)}{\bar{\sigma}_n}
\acontraction{\langle}{\sigma_n}{(u_1)\bar{\sigma}_n(v_1)\bar{\sigma}_n(u_2)}{\sigma_n}
\langle\sigma_n(u_1)\bar{\sigma}_n(v_1)\bar{\sigma}_n(u_2)\sigma_n(v_2)\rangle/(|u_1-v_2||v_1-u_2|)^{-\frac{c}{6}(n-\frac{1}{n})}\nn\\
&=\sum_p b_p \mathcal{F}(c,h_{\sigma_{n}}, h_p,1-x)\bar{\mathcal{F}}(c,\bar{h}_{\sigma_{n}}, \bar{h}_p,1-\bar{x}),
\end{align}
where $\mathcal{F}(c,h_{\sigma_{n}},h_p,x)$ and $\bar{\mathcal{F}}(c,\bar{h}_{\sigma_{n}},\bar{h}_p,\bar{x})$ are the Virasoro conformal blocks and $b_p$s are the OPE coefficients. 
We defined the cross ratio,
\be
x=\dfrac{(u_1-v_1)(u_2-v_2)}{(u_1-u_2)(v_1-v_2)},
\ee
and impose $x=\bar{x}$ since we are interested in the time slice $\tau=0$. The dominant contribution at the large-$c$ limit will come from a conformal family with the lowest scaling dimension in the channel\cite{Hartman:2013mia}. This approximation should be valid only for some specific region $x_c<x<1$. We do not specify the lower bound $x_c$, but just expect $x_c\sim\frac{1}{2}$. In this channel, the dominant one is universally $\sigma^2_n$ (and $\bar{\sigma}^2_n$) due to the twist number conservation,
\begin{align}
&\braket{\sigma_n(u_1)\bar{\sigma}_n(v_1)\bar{\sigma}_n(u_2)\sigma_n(v_2)}/(|u_1-v_2||v_1-u_2|)^{-\frac{c}{6}(n-\frac{1}{n})}\nn\\
&\sim b_{\sigma^2_n}\mathcal{F}(c,h_{\sigma_{n}},h_{\sigma^2_{n}},1-x)\bar{\mathcal{F}}(c,\bar{h}_{\sigma_{n}},\bar{h}_{\sigma^2_{n}},1-\bar{x}).
\end{align}
Next we would like to specify the analytic form of the above conformal blocks. 
This contribution of the conformal block consists only of light operators in the heavy-light limit\cite{Fitzpatrick:2014vua}. In this case, these analytic forms are known in the literature\cite{Fitzpatrick:2014vua, Hijano:2015zsa, Hijano:2015qja, Hijano:2015rla}. In our situation, the block for $\sigma^2_n$ has a simple form,
\be
\log\mathcal{F}(c,h_{\sigma_{n_o}},h_{\sigma^2_{n_o}},1-x)=-h_{\sigma_{n_o}}\log\left[\dfrac{1+\sqrt{x}}{1-\sqrt{x}}\right], 
\ee
where we assumed analytic continuation of odd integer $n\equiv n_o$ and the light limit $c\gg 1$ with fixed $h_{i}/c, h_{p}/c\ll 1$. Here we took the normalization in \cite{Hijano:2015zsa}. Therefore, we have obtained
\be
S_{o}(\rho_{A_1A_2})=S(\rho_{A_1A_2})+\dfrac{c}{6}\log\left[\dfrac{1+\sqrt{x}}{1-\sqrt{x}}\right]+\textrm{const.}\;,\label{eq:ewcs1}
\ee
where
\be
S(\rho_{A_1A_2})=\dfrac{c}{3}\log\dfrac{|u_1-v_2|}{\epsilon}+\dfrac{c}{3}\log\dfrac{|v_1-u_2|}{\epsilon}\;.
\ee
Here we introduced UV cutoff $\epsilon$. The constant terms do not depend on the position. For a while, we just assume the contribution from $b_{\sigma^2_n}$ is negligible at the large-$c$ limit. This assumption will be justified when we consider the pure state limit discussed in the next subsection. In the same way, we can compute the s-channel limit $x\rightarrow0$. In this case, the dominant contribution will be the vacuum block as like the EE. Hence, we obtain
\begin{align}
S_{o}(\rho_{A_1A_2})&=\dfrac{c}{3}\log\dfrac{|u_1-u_2|}{\epsilon}+\dfrac{c}{3}\log\dfrac{|v_1-v_2|}{\epsilon}\;\\
&=S(\rho_{A_1A_2}).\label{eq:ewcs2}
\end{align}
Therefore, we have confirmed
\be
\mathcal{E}_W(\rho_{A_1A_2})=\left\{
\begin{array}{ll}
\dfrac{1}{4G_N}\log\left[\dfrac{1+\sqrt{x}}{1-\sqrt{x}}\right] & (\textrm{t-channel}, x\sim1),\\
0 & (\textrm{s-channel},  x\sim0). \label{eq:ewcft2}
\end{array}\right.
\ee
in the two disjoint interval case. Here we used the relation between the central charge and the three-dimensional Newtonian constant $c=\frac{3}{2G_N}$\cite{Brown:1986nw}. The \eqref{eq:ewcft2} precisely matches the minimal entanglement wedge cross section for AdS${}_3$ (see FIG. \ref{fig:ewcs_single}). 
Extension to the multi-interval cases is straightforward.

\subsection{Pure state limit}
Let us consider the single interval limit $u_2\rightarrow v_1$ and $v_2\rightarrow u_1$. This corresponds to the pure state limit for the initial mixed state. In this case, our calculation reduces to a two point function of the twist operators. Hence, we can get the usual EE with single interval $A=[u_1,v_1]$ in this limit. This is generic statement for any CFT${}_2$, but let us see this behavior from \eqref{eq:ewcs1}. If one takes the distance $|u_2-v_1|$ and $|v_2-u_1|$ to the cutoff scale $\epsilon$, the second term of the right-hand side of \eqref{eq:ewcs1} reduces to the length of the geodesics anchored on the boundary points $u_1$ and $v_1$. Moreover, this argument guarantees the constant terms from $b_{\sigma_p^2}$ is irrelevant at the large-$c$ limit because of the position independence.
\section{Thermal state in holographic CFT${}_2$}\label{sec:4}
In this section, we consider the thermal state in holographic CFT${}_2$, which is genuinely mixed state and is dual to the static (planar) BTZ blackhole\cite{Banados:1992wn}. Namely, we will consider the CFT${}_2$ on cylinder $S^1_\beta\times\mathbb{R}$, with single interval on the time slice, $A=[-\ell/2,\ell/2]$. The $A^c$ denotes its complement. 

To compute $\textrm{Tr}_{\mathcal{H}}(\rho_{AA^c}^{T_{A^c}})^{n_o}$ by using the replica trick, one needs to take care about the location of the branch cut, which cannot be realized as the naive conformal map from the plane $z$ (previous results in section \ref{sec:3}) to the cylinder $w=\sigma+i\tau$. The correct prescription\cite{Calabrese:2014yza} is given by
\be
\textrm{Tr}(\rho_{AA^c}^{T_{A^c}})^{n_o}=\braket{\sigma_{n_o}(-L/2)\bar{\sigma}^2_{n_o}(-\ell/2)\sigma^2_{n_o}(\ell/2)\bar{\sigma}_{n_o}(L/2)}_\beta
\ee 
where we introduced a finite but large cutoff $L$ so that the conformal map can work. Thus, our ``complement'' $A^c$ is now $[-L/2,-\ell/2]\cup[\ell/2,L/2]$, although the true time slice is the infinite line. After taking the limit $n_o\rightarrow1$, we let $L\rightarrow\infty$ \cite{Calabrese:2014yza}. Here the suffix of correlation function $\beta$ denotes the inverse temperature. Then  the corresponding $\textrm{Tr}(\rho_{AA^c})^{n_o}$ should be
\be
\textrm{Tr}(\rho_{AA^c})^{n_o}=\braket{\sigma_{n_o}(-L/2)\bar{\sigma}_{n_o}(L/2)}_\beta
\ee
By using the conformal map $z=e^{2\pi w/\beta}$, one can write the above correlation function as
\begin{align}
&\textrm{Tr}(\rho_{AA^c}^{T_{A^c}})^{n_o}\nn\\
&=\left(\frac{2\pi}{\beta}\right)^{8h_{\sigma_{n_o}}}\braket{\sigma_{n_o}(e^{-\frac{\pi L}{\beta}})\bar{\sigma}^2_{n_o}(e^{-\frac{\pi \ell}{\beta}})\sigma^2_{n_o}(e^{\frac{\pi \ell}{\beta}})\bar{\sigma}_{n_o}(e^{\frac{\pi L}{\beta}})},\label{eq:thermal_ptn}\\
&\textrm{Tr}(\rho_{AA^c})^{n_o}
=\left(\frac{2\pi}{\beta}\right)^{4h_{\sigma_{n_o}}}\braket{\sigma_{n_o}(e^{-\frac{\pi L}{\beta}})\bar{\sigma}_{n_o}(e^{\frac{\pi L}{\beta}})}.
\end{align}
Then one can expand the \eqref{eq:thermal_ptn} by using the conformal blocks. The dominant contribution can be again approximated by the single conformal block contribution which depends on the value of the cross ratio. Here the cross ratio is $x=e^{-\frac{2\pi}{\beta}\ell}$ for sufficiently large $L$. 

\begin{figure}[htbp]
\begin{center}
\resizebox{45mm}{!}{\includegraphics{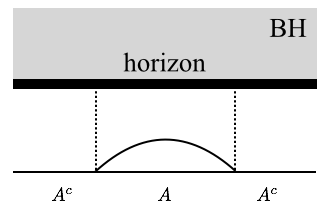}}
\caption{Calculation of $E_W$ for the static planar BTZ black hole. The inverse temperature $\beta$ is determined by the radius of the horizon. If the subsystem $A$ is sufficiently small $\ell\ll\beta$, the $E_W$ computes the geodesics anchored on the boundary of $A$ (black curve) which agrees with the \eqref{eq:ewcsbtz_x1}. For $\ell\gg\beta$, the $E_W$ does the disconnected surfaces (dotted vertical lines) which is consistent with the \eqref{eq:ewcsbtz_x0}.}\label{fig:btz}
\end{center}
\end{figure}

First, we consider the t-channel ($x\rightarrow1$) limit, $\ell\ll\beta$. Then  the dominant contribution from the channel is the vacuum block; hence, the \eqref{eq:thermal_ptn} reduces to the product of two point functions. After simple calculation, we obtain
\be
\mathcal{E}_W=\dfrac{c}{3}\log\dfrac{\beta}{\pi\e}\left(\sinh\dfrac{\pi\ell}{\beta}\right)+\textrm{const.}\;\;\; (x\sim1), \label{eq:ewcsbtz_x1}
\ee
where we introduced the UV cutoff $\epsilon$ form the dimensional analysis. The constant term comes from the normalization of two-point functions. This precisely matches the $E_W$ for the planar BTZ black hole (see FIG. \ref{fig:btz}). 

Next, we consider the s-channel ($x\rightarrow0$) limit, $\ell\gg\beta$. The dominant contribution in the channel is now the twist operator $\sigma_n (\bar{\sigma}_n)$ because of the twist number conservation. Then we have obtained
\be
\mathcal{E}_W=\dfrac{c}{3}\log\dfrac{\beta}{\pi\e}+\textrm{const.}\;\;\; (x\sim0),\label{eq:ewcsbtz_x0}
\ee
where the constant terms come from the normalization of two-point functions and the OPE coefficients. This again agrees with the $E_W$; however, it is important to note that this result is exact at the leading order of small $x$ expansion. There is the position-dependent deviation of order $\mathcal{O}(x^1)$. 

\section{Discussion}\label{sec:5}

The $\mathcal{E}_W$ can be {\it negative}. For example, one can find the Werner state in a 2-qubit system can have negative $\mathcal{E}_W$; thus, the $\mathcal{E}_W$ is farther from the entanglement measure than the EoP. The $\mathcal{E}_W(\rho_{AB})$ is rather similar to the coherent information $I(A\rangle B)\equiv S(\rho_B)-S(\rho_{AB})$\cite{Schumacher:1996dy, Horodecki:2005ehk}, or equivalently, the conditional entropy with the minus sign $S(A|B)\equiv-I(A\rangle B)$. Remarkably, these quantities can have both positive and negative values. The conditional entropy has already been discussed in the context of the differential entropy from which one can draw the bulk convex surfaces\cite{Balasubramanian:2013lsa,Myers:2014jia}. In particular, these were defined together with its orientation (with $\pm$ sign)\cite{Headrick:2014eia,Czech:2014ppa}. For the differential entropy, one needs infinite series of density matrices associated with each infinitesimal subregion. On the other hand, our present result has been derived from a single density matrix $\rho_{AB}$ dual to the entanglement wedge. This is a crucial difference compared with the differential entropy. It is very interesting to study the operational interpretation of $S_{o}$ as like the differential entropy\cite{Czech:2014tva}. Further studies on generic properties of the $S_o$ are also important.

From the viewpoint of the $\mathcal{E}_W$ in the present letter, the inequality\cite{Takayanagi:2017knl, Nguyen:2017yqw} $E_W(\rho_{AB})\geq I(A:B)/2$ is not always true and can impose new constraints on states dual to the classical geometries. Here we introduced the mutual information $I(A:B)=S(\rho_A)+S(\rho_B)-S(\rho_{AB})$. Understanding when such constraints can be satisfied is a very interesting future direction. It might be understood as the specific nature of the ``holographic states'' such as the absolutely maximally entangled (AME) states\cite{Pastawski:2015qua}. 
Derivation of the $E_W$ using the on-shell gravity action\cite{Faulkner:2013yia,Lewkowycz:2013nqa} would test our conjecture in general dimensions. Another obvious extension is to study the time-dependent setup on both sides. We would like to report on these issues in the near future. \vspace{-0.3cm}
\begin{acknowledgments}
We are grateful to Hayato Hirai, Norihiro Iizuka, Sotaro Sugishita, Tadashi Takayanagi, Satoshi Yamaguchi and Tsuyoshi Yokoya for useful discussion and interesting conversation. The author thanks Hayato Hirai, Norihiro Iizuka, Tadashi Takayanagi, and Satoshi Yamaguchi for useful comments on the draft.  
\end{acknowledgments}

\end{document}